\pdfoutput=1

\documentclass[twocolumn,showpacs,aps,prx,superscriptaddress]{revtex4-1}

\usepackage{graphicx}
\usepackage{upgreek}

\usepackage{amssymb}
\usepackage{amsmath}
\usepackage{physics}

\usepackage{color}
\usepackage{soul}

\usepackage{bm}
\renewcommand{\vec}[1]{\ensuremath{\bm{#1}}}

\renewcommand{\text}[1]{\ensuremath{\textrm{#1}}}
\newcommand{\ud}{\ensuremath{\text{d}}}

\DeclareMathOperator{\Si}{Si}
\DeclareMathOperator{\sinc}{sinc}
\DeclareMathOperator{\sign}{sign}

\begin{document}

\title{Bayesian inference for near-field interferometric tests of collapse models}

\author{Shaun Laing}
\author{James Bateman}
\email{j.e.bateman@swansea.ac.uk}
\affiliation{Department of Physics, College of Science, Swansea University, Swansea SA2 8PP, UK}

\date{\today}

\begin{abstract}
  We explore the information which proposed matterwave interferometry experiments with large test masses can provide about parameterizable extensions to quantum mechanics, such as have been proposed to explain the apparent quantum to classical transition. Specifically, we consider a matterwave near-field Talbot interferometer and Continuous Spontaneous Localisation (CSL). Using Bayesian inference we compute the effect of decoherence mechanisms including pressure and blackbody radiation, find estimates for the number of measurements required, and provide a procedure for optimal choice of experimental control variables. We show that in a simulated space based experiment it is possible to reach masses of $\sim10^9\,\text{u}$ and we quantify the bounds which can be placed on CSL. These specific results can be used to inform experimental design and the general approach can be applied to other parameterizable models.
\end{abstract}

\maketitle

\section{Introduction} \label{Sec:Intro}
Quantum mechanics is a remarkably successful physical theory for predicting microscopic behavior, successfully predicting and explaining phenomena such as the spectra of blackbody radiation~\cite{planck_theory_1914}, atom interferometry~\cite{kasevich_atomic_1991}, semi-conductors~\cite{shockley_electrons_1976}, and the properties of lasers~\cite{walls_quantum_2008} to name only a few.
There remains no experiment that falsifies or limits the regime of validity of quantum mechanics, and yet, at its core, there is an apparent problem where unitary evolution described by the Schr\"odinger equation gives way to a probabilistic description through the Born rule.
A thorough exploration of these problems and ideas can be found in Bassi el al.~\cite{bassi2013models}.

One intriguing possibility is that the apparent quantum-to-classical transition is real and there exists \emph{objective} spontaneous decoherence of the wavefunction, with the mechanism more effective for larger superpositions, with ``large'', in this context, interpreted to mean large in both mass and spatial separation~\cite{nimmrichter2013macroscopicity}.
Assessing this possibility is an experimental task.

\begin{figure}
  \includegraphics[width=0.9\columnwidth]{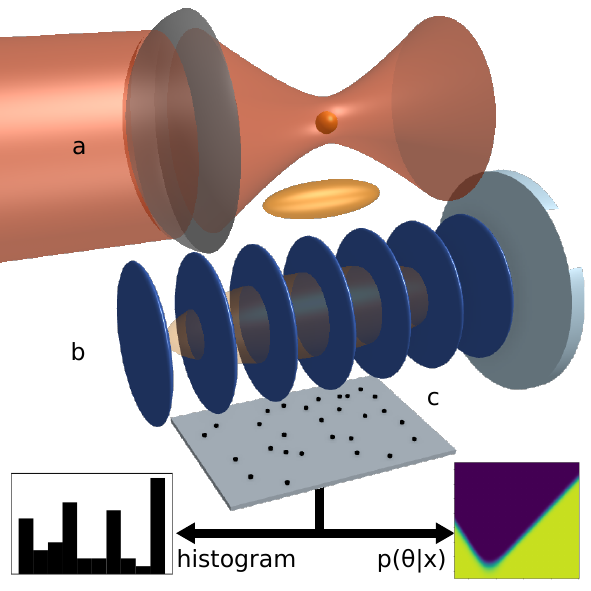}
  \caption{\label{Fig:1}Illustration of the scenario considered showing a particle localized in a harmonic dipole trap (a) the wavefunction of which, when released, expands to cover several fringes of a standing wave grating (b) formed by retroreflection of a light pulse.  The arrival location of particles is recorded (c) some time after the grating.  In typical scenarios, the number of data points is small and taking a histogram of arrival positions (left) \emph{may} be sufficient to evidence wave nature of the particle, the Bayesian inference approach (right; see main text) makes fuller use of the available information and can hope to constrain free parameters in CSL.}
\end{figure}
    
Experimental evidence which so far constrains these stochastic theories comprises both interferometric and non-interferometric measurements, the latter recording the (absence of) a predicted anomalous heating present in optomechanical systems. Such experiments include clamped cantilever systems using ferromagnetic spheres up to hundreds of ng~\cite{vinante_improved_2017}, tests for gravitational wave detectors using purely macrsocopic masses~\cite{carlesso_experimental_2016}, and levitated spheres in linear Paul traps~\cite{pontin_ultranarrow-linewidth_2020} or magneto-levitation traps~\cite{zheng_room_2020}.
Of interferometric experimental tests, the largest superpositions to-date come from Talbot--Lau matterwave interferometry with $25\,\text{ku}$ macromolecules~\cite{fein19_quant_super_molec_beyon_kda}.
These molecular beam-line experiments are remarkable, but to extend beyond this mass limit likely requires an  alternative approach.

Near-field Talbot effect applied to matterwave experiments with levitated nanoparticles~\cite{bateman2014nearfield} has emerged as a promising approach to interferometric experimental tests of stochastic theories~\cite{gasbarri21_prosp_near_field_inter_tests_collap_model}.
Creating such superpositions requires considerable effort to localize and measure position with sufficient accuracy and to mitigate sources of decoherence, including principally interactions with blackbody radiation and collisions with background gas.
This becomes extremely challenging for superposition which might inform the objective decoherence models as the timescale for free evolution---set by the Talbot time $t_\text{T} = md^2/h$ where $m$ is the mass, $d$ is the grating pitch, and $h$ is Planck's constant---can exceed tens of seconds.
Given such long free-evolution times, even with adequate position resolution and mitigation of decoherence, free-fall under gravity would mean that an Earth-bound experiment would be impractically tall.
Hence, space-borne experiments, where the particle remains nominally at rest relative to the apparatus, have been proposed and studied~\cite{kaltenbaek2016macroscopic,gasbarri2021testing,belenchia21_test_quant_mechan_space_inves_us_billion}, including a design study by the European Space Agency~\cite{esa2018qppf}.

The extent to which such an experiment would constrain particular theories is often expressed by excluding regions in parameter space representing the unknowns of the theory; most often this is Continuous Spontaneous Localization (CSL), which is parameterized by a rate $\lambda_c$ and a length-scale $r_c$~\cite{bassi2013models}.
Excluded regions are computed by noting that observation of an interference pattern with a given visibility would not be possible if there existed a given magnitude of objective collapse.
Predicted interference patterns may also be compared with classical predictions, perhaps with the aid of a figure of merit, to quantify the degree to which the proposed experiment necessitates a quantum description.
This approach does not assign a confidence interval nor does it compute the number of measurements necessary to exclude a region of parameter space.

In this manuscript we improve upon this method of excluding regions of parameter space by use of Bayesian inference which
assigns to these regions not a binary value of excluded or not, but a real-valued probability density.
We simulate the results of a hypothetical matter wave interferometry experiment and apply a Bayesian treatment to these results accounting for various sources of decoherence, such as blackbody radiation and collisions with gas molecules as is shown in Fig.~\ref{Fig:1}. This allows us to predict the bounds which such an experiment could reasonably set on the parameters of CSL.
In contrast with related work which considers tests of quantum mechanics more broadly, \cite{Schrinski2020}, we focus specifically on near-field matterwave interferometry, consider the two-dimensional parameter space of a specific collapse model with straightforward extensions to higher dimensions, and extend the description beyond the point-like particle approximation, as proves necessary for experimentally interesting scenarios.
This approach provides the ability to 
quantify the information gain which should be expected under given conditions,
compute the number of points necessary to reach a desired confidence,
and apply the extensive toolset of Bayesian optimal experiment design.
We are able to explore the experimental tradeoffs and data acquisition requirements that are acutely relevant for design of future experiments including a space-borne platform, such as a MAQRO like experiment~\cite{KaltenBaek2012}. We base our simulations on such an experiment, with parameters given in Table~\ref{table:Params}.

\section{Talbot Interferometer} \label{Sec:Talbot}

We consider a near-field Talbot interferometer consisting of a spherical nanoparticle of mass $m$, prepared in an thermal state of a harmonic trap of frequency $\Omega$, released and allowed to evolve freely for time $t_1$ before interaction with a phase grating with pitch $d$ and phase amplitude $\phi_0$, and finally free evolution for time $t_2$ before arrival position is measured~\cite{bateman2014nearfield}.  The initial thermal state has Gaussian position width $\sigma_x$ and momentum width $\sigma_p$.

The probability density for the particle arriving at position $x$ is given by
\begin{equation}
  \label{eq:TalbotPattern}
  \frac{m}{Z}\left[1+2\sum_{n=0}^\infty R_n B_n\left(\frac{ndt_2}{t_TD}\right)e^{-2\left(\frac{n\pi \sigma_x t_2}{Dt_1}\right)^2}\cos\left(\frac{2\pi nx}{D}\right)\right]
\end{equation}
where $Z=\sqrt{2\pi}\sigma_p(t_1+t_2)$,
$t_T=md^2/h$ is the Talbot time, $h$ is Planck's constant, and $D=d(t_1+t_2)/t_1$ is the geometrically magnified grating period to be expected at the detector~\cite{bateman2014nearfield,gasbarri21_prosp_near_field_inter_tests_collap_model}.

The functions $B_n$ are the generalized Talbot coefficients: they account for coherent and incoherent interaction of a spherical particle with a phase grating realized by a standing light wave from a pulsed laser~\cite{belenchia2019talbot-lau}. The terms $R_n$ describe decoherence and reduce the amplitude of each spatial frequency component: these account for the various sources of environmental decoherence, including blackbody radiation and collisions with gas particles, and the effect of any stochastic modification to quantum mechanics. 

Decoherence mechanisms, which we assume each act at a constant rate during the experiment~\cite[Sup.~Eq.~26]{bateman2014nearfield}, are characterised by a rate $\Gamma$ and a function $f(x)$ which characterizes the spatial extent of the localizing interaction.  The decoherence coefficients for each mechanism are computed as, 
\begin{equation} 
	R_n=\exp\left\{-\Gamma\left[1-f\left(\frac{nht_2}{mD}\right)\right](t_1+t_2)\right\}.
	\label{EQ:Deco_Full}
\end{equation}
A full treatment of the decoherence mechanisms is given in supplementary~\ref{Sup:Deco}.

Finite position resolution will impact any experiment and the measured position $x$ will be distributed about the true value.
We model this as a convolution with a Gaussian kernel of width $\sigma_m$, which may have a time dependence, and is described in supplementary~\ref{Sup:Pos_Res} for a MAQRO like experiment; via Fourier transforms and the convolution theorem we find that can be included as a decoherence term: $R_n^{\text{meas}} = \exp\left[-(2\pi n\sigma_m/D)^2/2\right]$.

Any experiment will operate over a finite spatial region $S$ and we include this via a multiplicative window,
\begin{equation}
	W(x) = 
	\begin{cases}
		1 & \text{if } x \in S\\
		0 & \text{if } x \notin S
	\end{cases}
	\label{EQ:Mult_Window}
\end{equation}

For brevity, we describe the pitch of the pattern using $k=2\pi/D$ and combine the pure Talbot terms as
\begin{equation}
	A_n=B_n\left(\frac{ndt_2}{t_TD}\right)\exp\left[{-2\left(\frac{n\pi \sigma_x t_2}{Dt_1}\right)^2}\right].
	\label{EQ:TalbTerms}
\end{equation}

\section{Bayesian model} \label{Sec:Bayes}
We consider any collapse model which manifests as a decoherence process and is described by a set of free parameters, which we write as vector $\vec{\theta}$.
The combined effect of several decoherence mechanisms is found by multiplying separate coefficients, and so we split the total decoherence into model and other contributions: $R_n=R_n^\text{mod}(\vec{\theta})R_n^\text{oth}$ where $R_n^{\text{mod}}(\vec{\theta})$ and $R_n^\text{oth}$ are respectively the effects of the decoherence from the objective collapse model and various environmental sources, including $R_n^\text{meas}$ described above.

Hence, rewriting Eq.~\ref{eq:TalbotPattern}, the joint probability density for \emph{measuring} position $x$ for specific values $\vec{\theta}$ is
\begin{equation}
  \label{eq:PxGivenTheta}
  p(x,\vec{\theta})= \frac{W(x) m}{Z}\left[1+2\sum_{n=0}^\infty R_n^\text{mod}(\vec{\theta})R_n^\text{oth}A_n\cos{\left(nkx\right)}\right].
\end{equation}
Using traditional rules of probability theory, we find the likelihood to be 
\begin{equation}
	p(x|\vec{\theta}) = 
	\frac{p(x,\vec{\theta})}
	{\int p(x,\vec{\theta}) dx}.
	\label{EQ:lkhd}
\end{equation}

Assuming independent and identically distributed position measurements, the joint probability of all $N$ measurements $\vec{x}=(x_1,x_2,\ldots x_N)$ is
$p(\vec{x}|\vec{\theta})=\Pi_{i=1}^Np(x_i|\vec{\theta})$.
Through Bayes theorem~\cite{sivia2006data}, this data $\vec{x}$ can be used to find the posterior probability for the parameters $\vec{\theta}$:
\begin{equation}
  \label{eq:Bayes}
  p(\vec{\theta}|\vec{x})\propto p(\vec{x}|\vec{\theta})p(\vec{\theta})
\end{equation}
where the constant of proportionality is $1/p(\vec{x})$, often called the `evidence', which we neglect here as it has no dependence on the parameters $\vec{\theta}$.

The prior $p(\vec{\theta})$ encapsulates our assumptions before considering any data, and must be approached with care.
Specifically, while e.g. uniform probability across $\vec{\theta}$ may seem natural, this is not invariant under parameterization i.e. our physical predictions would change if we wrote the model in a different form.
\par
The canonical choice for an uninformative prior is Jeffrey's prior, proportional to the square root of the determinant of the Fisher information matrix.  However, this prior does not extend well into multi-dimensional problems~\cite[\S5.2.9]{lunn_prior_2013} giving unphysical posteriors for the collapse model which we consider here.

We choose the Maximal Data Information Prior (MDIP) which, while derived from the likelihood of a measurement and therefore dependent on the choice of parameterization, minimizes the arbitrary choices and is designed to maximize the information gain from a single measurement~\cite{zellner_introduction_1996}. Further, it has no difficulty working with problems of arbitrary dimension~\cite{zellner_models_1996}. It is given by~\cite{yang_catalog_1998},
\begin{equation}
	p(\vec{\theta}) \propto \exp [\int p(x|\vec{\theta})\ln p(x|\vec{\theta})dx].
	\label{eq:MDIPDef}
\end{equation}

Design of this prior ensures that even for a small number of measurements the posterior quickly becomes dominated by updates from the data.  The prior is approximately flat but does inherit some shape from the experimental design, reminiscent of plots from the literature on experiments which seek to constrain parameters of CSL.

The MDIP makes no reference to other experiments which provide considerable information on what values of CSL are plausible. Therefore we also consider a prior, which we term the ``Experimental Prior", informed by the results of all previous experiments. We define this Experiential Prior as constant over all values of $\vec{\theta}$ that have not yet been excluded by experiment and zero for those values which have been excluded by previous experiments~\cite{carlesso_present_2022}. 

\section{Continuous Spontaneous Localization} \label{Sec:CSL}

Until now our discussions have been for general collapse models. From now we focus on the model of Continuous Spontaneous Localization (CSL)~\cite{bassi2013models}. CSL can be described as a decoherence term of the form of Eq.~\eqref{EQ:Deco_Full} defined by its rate and length scale parameters which are to be estimated empirically. The decoherence effects of CSL enter the total decoherence of the interferometer as $R_n^\text{mod}(\vec{\theta})$ where $\vec{\theta} = [\lambda_c,r_c]$ is a vector containing the CSL parameters.
The description was recently extended to include non-point-like particles~\cite{gasbarri21_prosp_near_field_inter_tests_collap_model}.  Using a spherical test particle of radius $R$, mass $m$, and constant density $\rho$ in $0\leq r\leq R$, we have dimensionless pre-factor $A=\left(36/\sqrt{\pi}\right)\left(m/m_0\right)^2\left(r_c/R\right)^2$ and 
\begin{align}
  \Gamma_\text{CSL}
  &=A\,\lambda_c\,
  \int_0^\infty e^{-\alpha^2} j_1\left(\frac{\alpha R}{r_C}\right)^2\ud\alpha,\\
  f(x)
  &=\frac{A}{\Gamma_\text{CSL}}\,\lambda_c\,\frac{r_c}{x}
  \int_0^\infty e^{-\alpha^2} j_1\left(\frac{\alpha R}{r_C}\right)^2 \frac{1}{\alpha}\Si\left(\frac{\alpha x}{r_C}\right)\, \ud \alpha
\end{align}
where $j_1$ is the Spherical Bessel function of the first kind, $\Si$ is the sine integral, and $m_0$ is the atomic mass constant.  These arise via the Fourier transform of the uniform density sphere $\tilde{\mu}(q)=4\pi\hbar R^2 \rho\, j_1(qR/\hbar)/q$; further details in Supplementary~\ref{Sup:CSL}.

\begin{figure*}
\includegraphics[width=\linewidth]{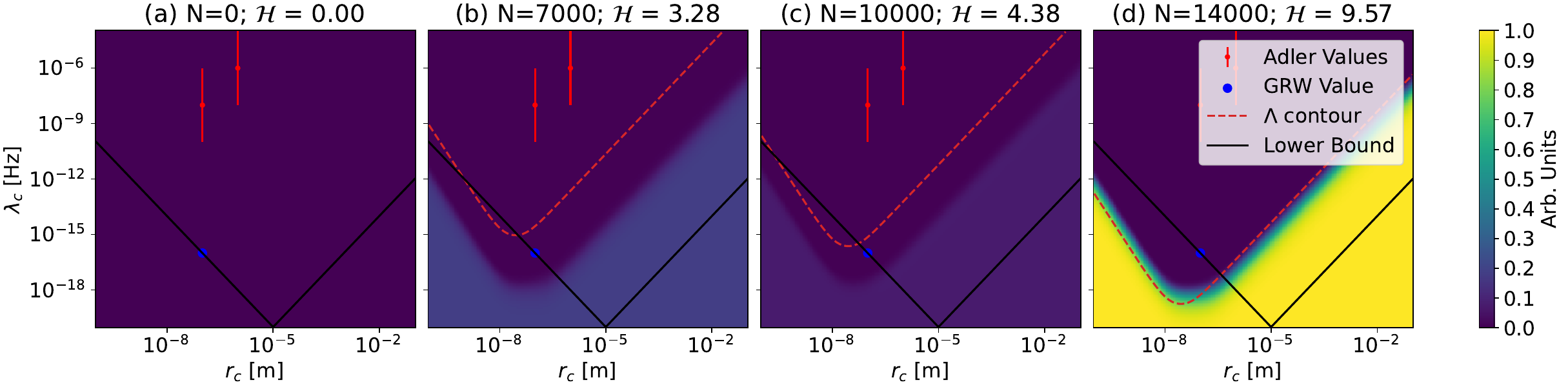}
\caption{\label{fig:posteriors} Probability distributions $p(\vec{\theta}|\vec{x})$ for the rate ($\lambda_c$) and length scale ($r_c$) parameters of the CSL model with MAQRO like parameters, as detailed in the text. These posteriors are generated using the Maximal Data Information Prior, with $N$ points picked from distribution $p(x|\vec{\theta}=0)$. (a) $N=0$ is the prior; (b--d) are for $N$ as indicated.  The GRW value \cite{ghirardi_unified_1986}, Adler values \cite{adler_lower_2007}, and lower bound, discussed in the text, are motivated by theoretical considerations and are only shown for comparison.
The red dashed upper bound is found such that $95\%$ of the probability distribution is below this line.}
\end{figure*}

\section{Posterior probabilities} \label{Sec:Posts}

Our discussion now focuses on application of Bayesian inference for Talbot interference experiments in the specific case of the CSL model.
We compute the expectation value of the posterior distribution under the hypothesis that $\vec{\theta}=0$, i.e. that the CSL effect, if it exists, is vanishingly small on the scale accessible to the experiment.
For a more general discussion on hypothesis falsification in the context of macroscopic superpositions, see~\cite{schrinski2019macroscopicity}.

\begin{table}
	\centering
	\begin{tabular}{c c c} 
 		\hline \hline
 		Symbol & Name & Value \\ [0.5ex] 
 		\hline
 		$\rho$ & Si particle density & $2329 \text{kg m}^{-3}$ \\ 
 		$\lambda_G$ & Grating laser wavelength & $2d = 354\text{nm}$ \\
 		$\Omega$ & Trapping frequency & $200 \text{kHz}$ \\ 
 		$T_{\text{com}}$ & Initial center of mass temperature & $20\text{mK}$ \\
 		$T_{\text{int}}$ & Initial internal temperature & $25\text{K}$\\ 
 		$T_{\text{env}}$ & Environmental temperature & $20\text{K}$ \\
 		$t_1$ & Pre-grating time & $2t_T$\\ 
 		$P_g$ & Residual gas pressure & $10^{-15}\text{hPa}$\\
 		$\sigma_x$ & Gaussian Position Width & $\sqrt{k_B T_\text{com}/4\pi^2 m \Omega^2}$\\
 		$\sigma_m$ & Measurement Position Uncertainty& $\sigma_x + (10\text{nm}/100\text{s})t$\\[1ex] 
 		\hline\hline

	\end{tabular}
	\caption{Control parameters used in the MAQRO-like scenario. The values of the free-fall time $t_2$, and phase parameter $\phi_0$ are optimized for each new run based on the method described above.}
	\label{table:Params}
\end{table}

Fig.~\ref{fig:posteriors} shows prior and posterior distributions, for simulated measurements of a silicon nanoparticle with a mass of $m = 10^8\,\text{u}$ for parameters typical of a MAQRO like experiment, as given in Table~\ref{table:Params}. The phase parameter $\phi_0$ and second free-fall time $t_2$ are chosen via the optimization process described in Sec.~\ref{Sec:Opt_Params} to maximize the change in visibility brought on by the collapse model.
Sources of decoherence used the simulations are discussed in Supplementary~\ref{Sup:Deco}.

The posterior distributions in Fig.~\ref{fig:posteriors} are each from a single realization of $N$ simulated arrival positions. While in principle we should compute the expectation value over many realizations, the distributions are indistinguishable for $N\gtrsim 1000$.

We indicate on the posteriors values suggested by Adler~\cite{adler_lower_2007} based on the rate of latent image formation in photography or etched track detection, and the value suggested by Ghirardi, Rimini, and Weber~\cite{ghirardi_unified_1986} that was chosen to ensure that quantum systems maintained their coherence for times comparable to the age of the universe while macroscopic systems collapse in fractions of a second. The black line in the plots indicates the lower bound on the CSL parameters; it is derived not from experiments but from the requirement that macroscopic superpositions do not maintain their coherence for long periods of time. The specific values given are chosen such that a graphene disk with radius $10\upmu \text{m}$ collapses in $0.01\text{s}$. This ensures that the smallest sized objects that can be detected by the human eye collapses in the time resolution of the eye~\cite{carlesso_present_2022}. It should be noted that the Experimental Prior we consider does not consider these lower limits, remaining non-zero for all values bellow the empirically determined upper bounds.

The plot region is finite in extent and contains much of the probability space of interest. The logarithmic scale means the integrated region of lower $\lambda_c$ and $r_c$ is small, and the distribution decays quickly in the region of larger $\lambda_c$ and $r_c$; the probability value is small for regions of larger $\lambda_c$ and $r_c$ which have also been previously excluded by non-interferometric experiments.

The posteriors include an upper bound line for the possible values of the CSL parameters $\lambda_c$ and $r_c$. This curve is dependent on the value of the decoherence strength $\Lambda$ which governs the rate at which each spatial frequency is reduced by a given source of decoherence. The relationship between the CSL parameters and the decoherence strength is given in Supplementary~\ref{Sup:Exclusion}. To define the location of the upper bounds, we adjust the value of $\Lambda$ via iterative bisection until the integral of the posterior below the line is approximately $95\%$ of the total, i.e. there is a $95\%$ probability that, based on measurements, the true value falls below this line.

As we collect data about particle arrival positions a plausible region of $\vec{\theta}$ begins to emerge leaving behind an upper triangle of classicality. We begin to see the difference between regions after $N\sim 4\times 10^3$ and the difference between these regions only increases as $N$ increases. Beyond $N\sim 10^4$ we reach an asymptotic limit where acquisition of more data does not appreciably change the distribution.
Quantification of these statements, and comparison across different scenarios, is the subject of the next sections.

\section{Information Gain}\label{Sec:Information}

Information theory provides us with an objective means to quantify the information gain from a given experiment.  An objective measure of the information about CSL provided by a measurement is essential to make informed decisions about the number of measurements we should plan to take and, more broadly, the optimal design of an experiment.

The Kullback--Leibler divergence offers a measure of the information contained within the posterior probability through comparison of this with the prior~\cite{kullback78_infor,bishop06_patter,mackay03_infor,sivia2006data}.  The information contained in data $\vec{x}$ is hence
\begin{equation}\label{eq:information}
  \mathcal{H}(\vec{x})=\int p(\vec{\theta}|\vec{x})\log_2\left[\frac{p(\vec{\theta}|\vec{x})}{p(\vec{\theta})}\right]\ud\vec\theta.
\end{equation}

We can compute the \emph{expected information} for a given experiment by integrating over all possible outcomes, weighted by the probability of each outcome occurring:
\begin{equation}
  \langle\mathcal{H}\rangle = \int \mathcal{H}(\vec{x}) p(\vec{x}) \ud\vec{x}
  \label{eq:expectedInformation}
\end{equation}
where $p(\vec{x})$ is the `evidence', seen in Eq.~\ref{eq:Bayes} where it was treated as a proportionality constant.  It can be obtained as
\begin{equation}
  p(\vec{x})=\int p(\vec{x}|\vec{\theta})p(\vec{\theta})\ud\vec{\theta}.
  \label{eq:evidence}
\end{equation}

Expected information Eq.~\ref{eq:expectedInformation} is found by integration over the space of all possible experimental outcomes $\vec{x}$.
Due to the large number of dimensions in $\vec{x}$, direct integration is numerically intractable. 
As a result we use the Monte-Carlo integration method described in \cite{Huan2013}. 
We rewrite Eq.~\eqref{eq:expectedInformation} as, 
\begin{equation}
	\begin{split}
	 	\langle\mathcal{H}\rangle &= \iint \{\ln[p(\vec{x}|\vec{\theta})] - \ln[p(\vec{x})]\}p(\vec{x}|\vec{\theta})p(\vec{\theta}) d\vec{\theta} d\vec{x}\\
	 	&\approx \frac{1}{M} \sum_{i=1}^M \{\ln[p(\vec{x}^{(i)}|\vec{\theta}^{(i)})] - \ln[p(\vec{x}^{(i)})]\}
	 \end{split}
	 \label{eq:Huan_MCMC}
\end{equation}
where the values of $\vec{\theta}^{(i)}$ are drawn from the prior $p(\vec{\theta})$ and the data $\vec{x}^{(i)}$ is drawn from the conditional likelihood $p(\vec{x}|\vec{\theta} = \vec{\theta}^{(i)})$.
Although there is no analytical expression for the evidence $p(\vec{x}^{(i)})$ in this calculation, due to the low dimensionality of $\vec{\theta}$, we are able to calculate it via numerical integration~\cite{Ryan2003}.

We estimate the uncertainty of $\langle \mathcal{H}\rangle$ using statistical error, $\Delta = \sqrt{(\langle \mathcal{H}^2 \rangle - \langle \mathcal{H} \rangle^2)/M}$ where $M$ is the number of Monte--Carlo iterations performed. These bounds are shown in our plots as the shaded regions around the mean values.

\begin{figure}
\includegraphics[width=\linewidth]{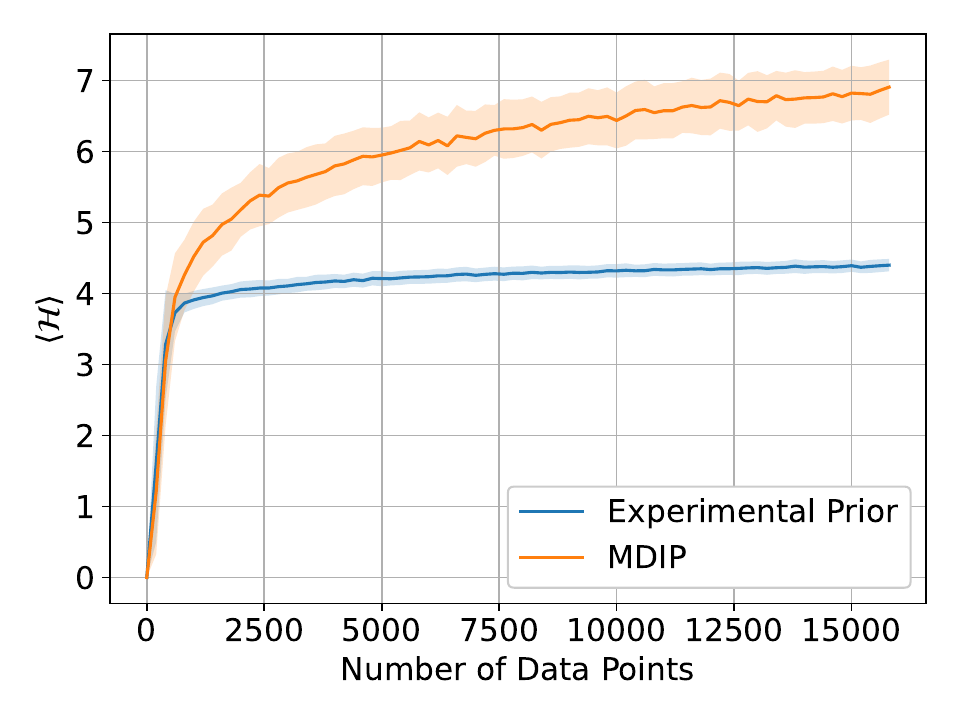}
\caption{\label{fig:expectHvsN}Expected information $\langle \mathcal{H}\rangle$ as a function of the number of data points $N$ for a MAQRO like scenario as discussed in the text with a nanoparticle of mass $m = 10^{8}\text{u}$ for setting $\vec{\theta}=0$. The shaded region around each line indicates the standard deviation in the estimation. The information gain is small beyound $N\sim 10^4$ and the variance in our estimation of this value changes only very slowly.  These are computationally expensive and for a given $N$ the value of $\langle H\rangle$ is found from 200 realisations of $H\vec(x)$.}
\end{figure}

Fig.~\ref{fig:expectHvsN} shows the expected information gain under the assumption CSL is false, i.e. the measured data is distributed about $p(x_i|\vec{\theta}=0)$, as a function of the number of data points $N$ for the MAQRO scenario and using the Maximal Information Data Prior and the Experimental Prior.  This quantifies earlier statements about the asymptotic behavior of the distribution; we see that the information gain per data point falls to negligible values beyond $N\sim 10^4$. We can also see the difference in the amount of information gained per data point for different priors.

\begin{figure}
\includegraphics[width=\linewidth]{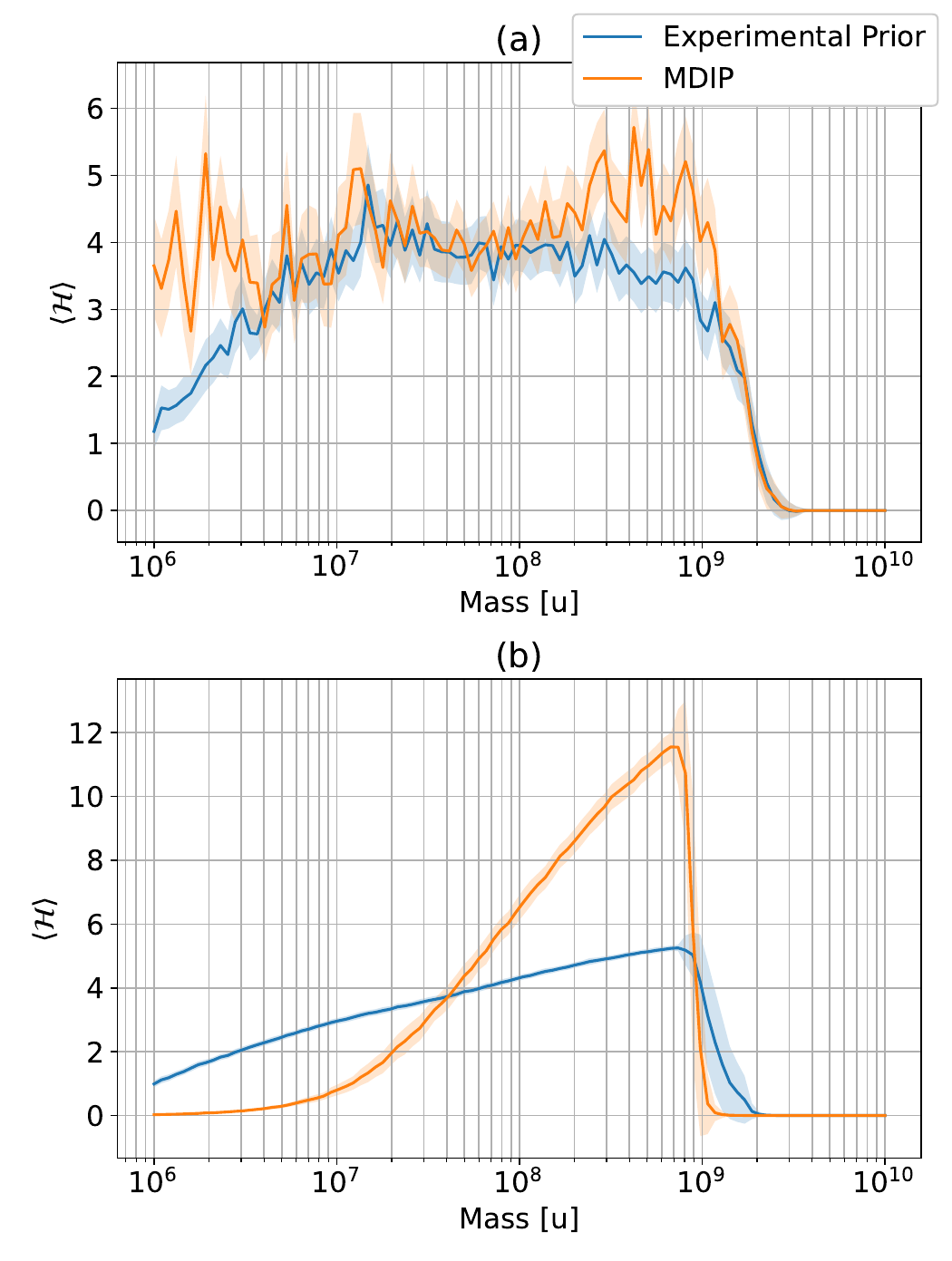}
\caption{Maximum expected information for different priors as a function of particle mass. The orange lines show the information gained from the MDIP, which is mass dependent, and the blue lines show the information gained from using a prior motivated by previous experimental results. (a) show the expected information found through Markov Chain Monte-Carlo integration distributing our measurements by $p(\vec{x})$ as described in Eq.~\eqref{eq:Huan_MCMC} ,while (b) shows the expected information under the assumption that there is no CSL effect, i.e. $\vec{\theta} = 0$.}
\label{fig:InfoAve}
\end{figure}

The expected information gain after $10^4$ measurements for two different priors is shown in Fig. \ref{fig:InfoAve}. Panel (a) shows the expected information gain using the Markov Chain method described in Eq. \eqref{eq:Huan_MCMC} whilst panel (b) shows the information we would gain if there is no CSL effect. 
We see that the choice in prior has a large impact on the information gain we compute. 
This is because the information of Eq. \eqref{eq:information} gives the amount of information contained in the posterior relative to the prior, and the MDIP is designed to be least informative. We see in the simulations using the MDIP that the MAQRO like scenario (given in Table \ref{table:Params}) achieves a maximum information near $M\sim 10^9\,\text{u}$. At this point the radius of the silicon nanoparticle used is comparable to the grating width and thus to achieve an appreciable phase-shift requires a large fluence for the pulsed laser enacting the grating, leading inevitably to large decoherence from this interaction.

We note that by using a non-invariant prior we become bound by the choice of parameterization which is, in this case linear i.e. $p(\vec{x}|\vec{\theta})=[\lambda_c,r_c]$.

\begin{figure}
	\centering
	\includegraphics[width=\linewidth]{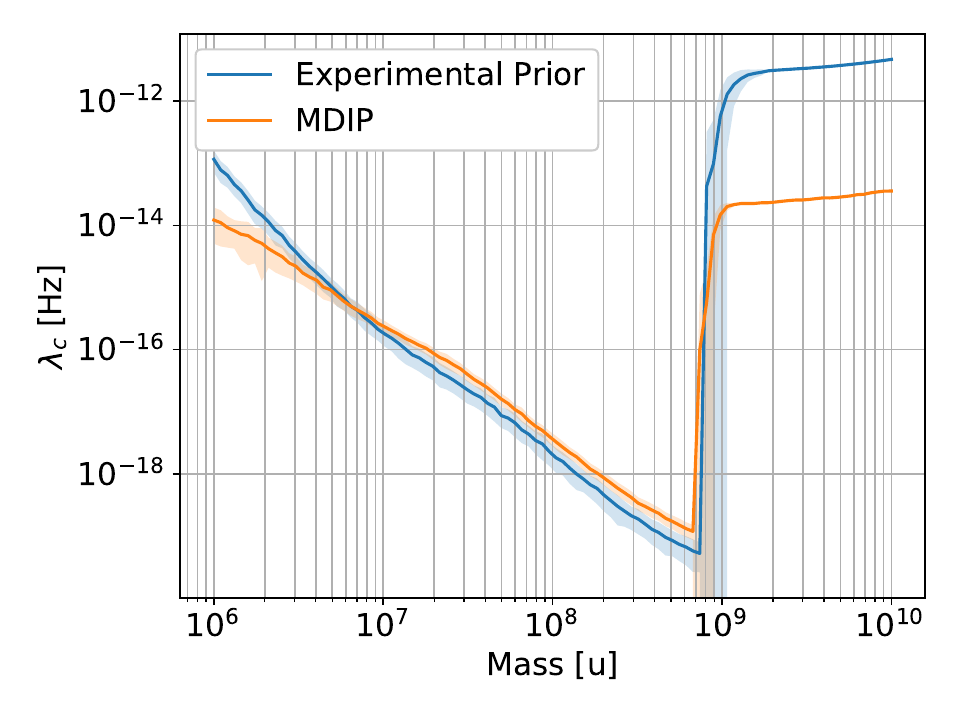}
	\caption{The lowest upper bound that can be put on the value of $\lambda_c$ at $r_c = 10^{-7} \text{m}$ as a function of particle mass. The value of the upper bounds is found from the posterior after $10^4$ measurements starting from either the MDIP (orange line) or an experimentally motivated posterior (blue line). We note the difference in the parameter $\lambda_c$ for large masses (~$10^9\text{u}$) occurring from the fact that regions of the experimental prior are set to identically 0, such that the 95\% confidence line is found to be lower than for posteriors generated by the MDIP.}
	\label{fig:Lambda_plot}
\end{figure}

Previous works have characterized the capabilities of matterwave experiments by a single decoherence parameter $\lambda_c$ when setting $r_c = 10^{-7} \text{m}$~\cite{bateman2014nearfield,nimmrichter_testing_2011}, which allows for simple comparison with the GRW and Adler suggested values for the CSL parameters. Hence, we compute the value of $\lambda_c$ at this choice of $r_c$ to facilitate comparison with our measure $\langle\mathcal{H}\rangle$. Fig.~\ref{fig:Lambda_plot} shows that the expected value of $\lambda_c$ given $\vec{\theta} = 0$ decreases as the particle mass is increased. This continues until the particle reaches a mass of $~10^9 \text{u}$ where the information falls to $0$. At this point the value of $\lambda_c$ rises significantly. This value at high masses is an artefact of the finite integration region of the parameter space we are considering.
The information gain provides a more general metric as it makes use of the full parameter space.

\subsection{Optimal Experimental Design} \label{Sec:Opt_Params}

Our experiment is parameterized by a set of control parameters $\vec{C}$. This is a vector containing all the parameters that we can, in principle, choose in advance of performing the experiment, consisting of the particle mass $m$, the free fall times $t_1$ and $t_2$, the phase parameter $\phi_0$. We seek values for the parameters $\vec{C}$ which maximize the amount of information gain expected in each measurement. In principle we would choose $\vec{C}$ so as to maximize the value of $\langle \mathcal{H} \rangle$. However, this is computationally expensive, so as a proxy, we find $\vec{C}$ such that the introduction of CSL has a maximal effect on the expected fringe visibility.

We define visibility as the first order term from Eq.~\eqref{eq:TalbotPattern},
\begin{equation}
	\nu_\text{sin}(\vec{C}) = 2\beta \left|B_1\left(\frac{dt_2}{t_TD},\vec{C}\right)\right|
	\exp\left[-2\left(\frac{\pi \sigma_x t_2}{Dt_1}\right)^2\right]
	\label{eq:sin_vis}
\end{equation}
which captures the maximum amplitude of the interference pattern without any decoherence effects. We also define the visibility after the CSL effect as,
\begin{equation}
	\nu_\text{red}(\vec{C}) = \nu_\text{sin}(\vec{C})R_1^\text{mod}(\vec{\theta},\vec{C}).
\end{equation}

We then find $\vec{C}$ such that we maximize the difference $\nu_{sin}(\vec{C})-\nu_{red}(\vec{C})$. The optimum values of the parameters $\phi_0$ and $t_2$ are shown for various particle masses in Fig.~\ref{fig:Opt_Params} in the supplementary.
	
For the purpose of this study we consider the mass $m$ to be given parameter and not subject to experimental control.  This is a simplification because, in a real experiment, we anticipate that $m$ will be measured in-situ for each particle.  Further, we assume that time before the grating, $t_1$, is also given, as this depends on mass and the grating pitch, and does not crucially affect the interference pattern provided it is sufficiently long for spatial coherence to emerge.  Hence, we focus our study of control parameters on $\vec{C}=[t_2,\phi_0]$.
\par
Using the method of parameter optimization described above, we plot the maximum expected information $\langle \mathcal{H}\rangle_\infty$ in the limit of a large number of measurements $N$, for an experiment using a particle of a given mass as shown in Fig.~\ref{fig:InfoAve}. As shown in Fig.~\ref{fig:expectHvsN}, after $N \approx 10^4$ the information gained per new data point becomes small, so we choose the limit of large $N$ to be $N = 10^4$. Armed with these tools, we proceed to compare the ultimate limits of expected information under different experimental conditions.

\section{Scenario comparison} \label{Sec:Scenarios}

The notion of expected information facilitates comparison across difference scenarios.  The MDIP described earlier inherits structure from the experimental design. However, due to the mostly flat nature of the prior we can use it to compare the information gain between experiments.

\begin{figure}
\includegraphics[width=\linewidth]{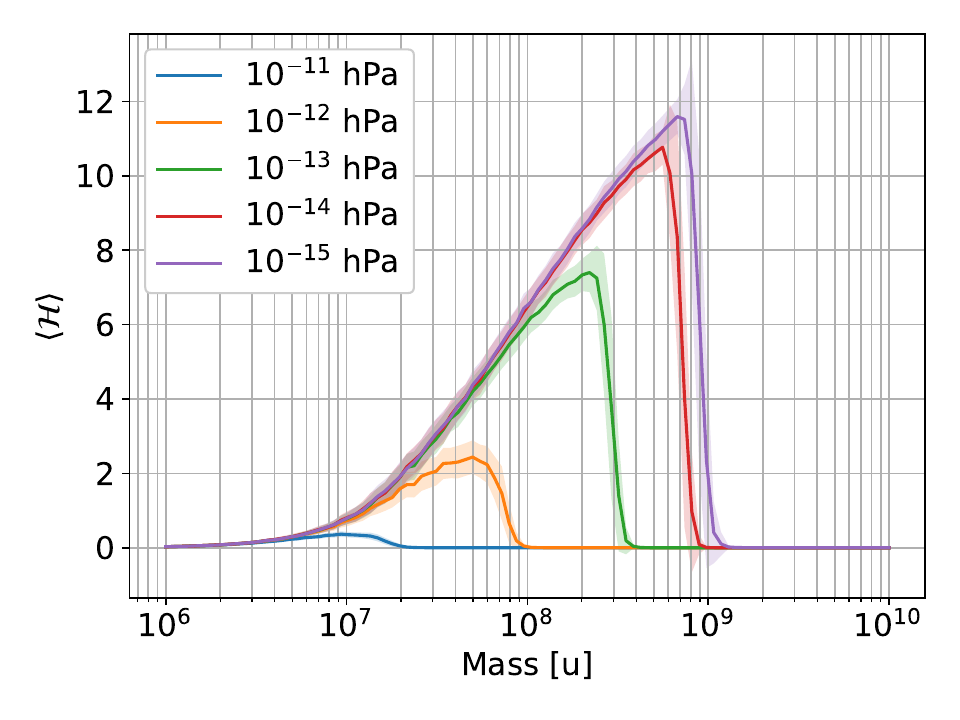}
\caption{The expected information under the assumption that there is no measurable CSL effect for various environmental pressures as a function of particle mass for a range of scenarios in a MAQRO like experiment.}
\label{fig:expectHvsM}
\end{figure}

Fig.~\ref{fig:expectHvsM} shows the expected information gain given $\vec{\theta}=0$ for various environmental pressures. It shows that in order to maximize the mass of the test particle, we must decrease the pressure, because for higher pressures the decoherence is dominated by collisions with residual gas particles. While this general point is well known, we are now able to quantify this pressure. We observe that pressures below $10^{-14}\,\text{hPa}$ are sufficient to maximise the particle mass and information we can gain (see Figs.~\ref{fig:DecoSources} and~\ref{fig:TotDeco} in the supplementary).

\begin{figure}
	\centering
	\includegraphics[width=0.5\textwidth]{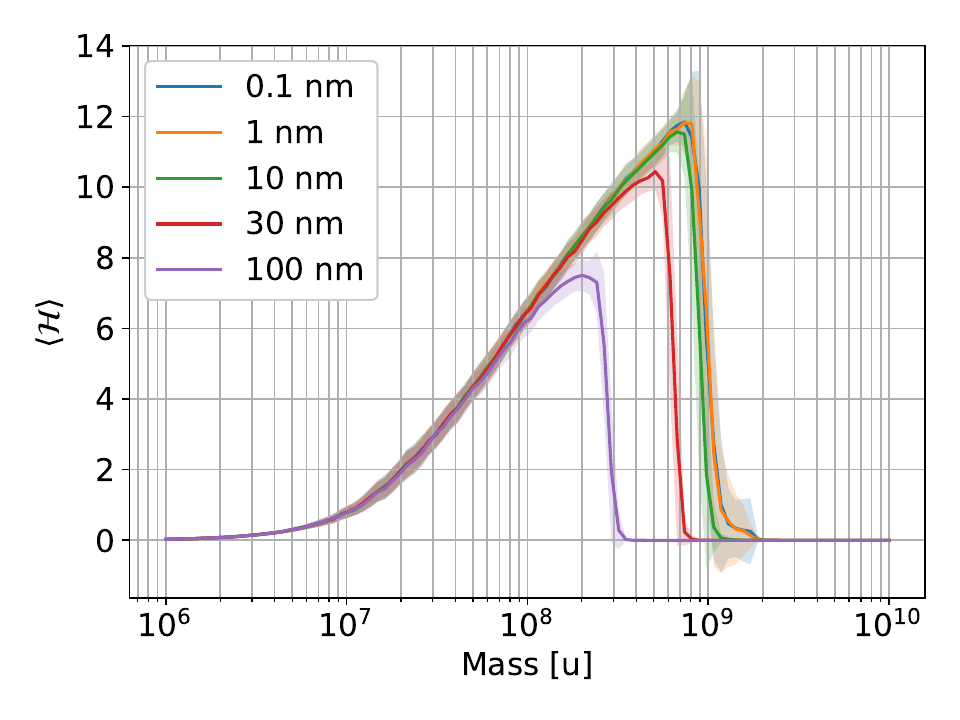}
	\caption{The expected information under the assumption that there is no measurable CSL effect for various drift uncertainty parameters as a function of particle mass for a range of scenarios in a MAQRO like experiment. Each new plot is found by changing the width per $100 \text{s}$ of increase in the position uncertainty.}
	\label{FIG:Drift_Uncert}
\end{figure}

A further consideration is the maximum allowable measurment uncertainty ans in many scenarios this increases linearly with free flight time. Fig.~\ref{FIG:Drift_Uncert} shows the expected information gain given $\vec{\theta}=0$ for various drift uncertainty parameters $\sigma_m$ given in Table~\ref{table:Params}. Minimising the increase in position uncertainty allows us to gain information from higher masses. However, for drifts better than $10 \text{nm}$ in $100 \text{s}$, as was suggested by the ESA design study \cite{esa2018qppf}, there is negligible effect on expected information for MAQRO like experiments.

\section{Conclusions and future work}

We have presented a procedure and computed the probability density which a MAQRO like experiment is likely to assign to parameters of CSL as a popular example of a parameterizable macrorealistic extension to quantum mechanics.
This work provides a toolbox for exploring specific scenarios as part of a design study. We show that in a MAQRO like experiment, around $10^4$ measured arrival locations are needed to saturate the information gain. We also find that environmental pressures below $10^{-14} \text{hPa}$ are sufficient to maximise the particle mass with which we can achieve superpositions. This is in contrast to previous proposals that suggest tolerances to much higher pressures, and we provide for the consequences if this stringent pressure requirement cannot be met.
Finally we show that increases in the uncertainty due to spacecraft drift up to $10 \text{nm}$ in every $100 \text{s}$ maximise the range of particle masses that can achieve superpositions.

Future work will include optimal experimental design where we provide a strategy such that the optimal parameters can be chosen based on all previous data $\vec{x}$.

The current description and approach of MAQRO is to minimize known sources of decoherence so that the effect of CSL can be observed.
Any claim of observing non-zero CSL parameters will then be dependent on confident knowledge of these decoherence sources.
An improved approach would be to distinguish the effect of CSL from other sources by its dependence on parameters such as mass and free-flight time.
Future work will apply the Bayesian description herein to define a strategy which can infer parameters of CSL in the presence of imperfectly known sources of decoherence.

\section{Acknowledgements}
We would like to thank G. Gasbarri for helpful discussions and suggestions.
S. Laing is grateful for EPSRC support through Standard Research Studentship (DTP) EP/R51312X/1.

J.B. conceived of the presented idea. S.L. developed the theory and computational simulations with J.B. providing contributions. All authors discussed the results and contributed to the final manuscript.

\bibliography{bibliography}

\appendix

\section{Numerical implementation} \label{Sup:Numeric}

The numerical implementation leverages the capabilities of Python and the NumPy~\cite{harris20_array_progr_with_numpy} and SciPy~\cite{virtanen20_scipy} libraries.  Details of the implementation of several key equations is given in the following sections.

\subsection{Window and sampling} \label{Sup:Window}

As discussed in the main text, it is necessary to include a finite window $W(x)$ so that the distribution is normalizable.  This window must cover a large number of oscillations, and the sampling must be sufficiently fine that oscillations are well resolved. The grating pitch is $100\,\text{nm}$ which is scaled geometrically by $(t_1+t_2)/t_2$ to, typically, $180\,\text{nm}$.  We include terms up to 6th order, and so, by Nyquist, must sample once every $10\,\text{nm}$.  We chose a $10\,\upmu\text{m}$ square-edged window which covers $\sim 50$ complete oscillations which we sample with 1000 points.

\subsection{Grating phase} \label{Sup:Grating}

Grating phase depends on the grating laser focus area $a_G$, the pulse energy $E_G$, and the properties of the nanoparticle.  For a given mass, we chose pulse area such that it covers the thermal distribution after expansion for time $t_1$, and we then chose the pulse energy such that $\phi_0 = \frac{4F_0}{\hbar ck^3} \frac{E_G}{a_G}$~\cite{belenchia2019talbot-lau} matches the optimum value according to the visibility argument described in the main text.

\subsection{Optimum parameters} \label{Sup:Opt_Params}

\begin{figure}
	\includegraphics[width=\linewidth]{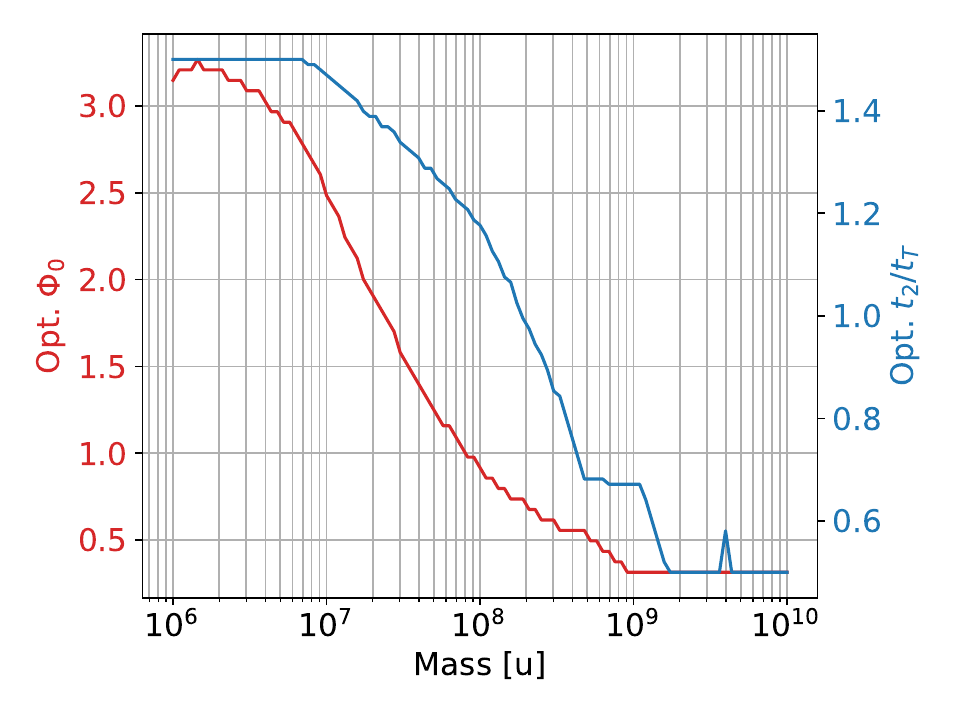}
	\caption{Optimum values of $\phi_0$ and $t_2$ for various masses of nanosphere. The bounds on $t_2$ are chosen to ensure that $t_2 \approx t_T$ to avoid excessive flight times increasing decoherence whilst allowing enough time for the interference effect to take place.}
	\label{fig:Opt_Params}
\end{figure}

The specific experiment is defined by the set of control parameters $\vec{C}$ as discussed in the main text. For many of the parameters, we can use physical arguments to set their values. For example the residual gas pressure $P_g$ should be as low as possible to minimize decoherence due to gas collisions. However, the parameters $\phi_0$ and $t_2$ do not have such obvious values, so we follow the method given in the main text to chose the optimum values for these parameters. The optimum values for these parameters are given for various particle masses in Fig.~\ref{fig:Opt_Params}.

\subsection{Determining Maximum Exclusion Line} \label{Sup:Exclusion}

We use the decoherence strength $\Lambda$ to describe how much of the parameter space has been excluded by a specific experiment. We describe the reduction effects on the fringe visibility as,
\begin{equation}
	R_n = \exp \left[-\frac{\Lambda (t_1+t_2) (n\kappa d)2}{3}\right]
	\label{eq:LambdaReduc}
\end{equation}
where $\kappa = \frac{t_1t_2}{(t_1+t_2)t_T}$ and $\Lambda$ is the parameter governing the strength of the decoherence~\cite{kaltenbaek2016macroscopic}. By setting $n=1$ to only consider the visibility we can compare this with the CSL reduction term to find $\Lambda$ as a function of $\lambda_c$ and $r_c$,
\begin{equation}
	\Lambda = \frac{-3\Gamma [f(x) - 1]}{(\kappa d)^2},
	\label{eq:LamDec}
\end{equation}
where $f(x)$ is the resolution factor of the collapse mechanism. It is given by Eq,~\eqref{EQ:CSLf(x)} for CSL.
This parameter can be used to define the upper exclusion line for the CSL parameters. We simply rearrange Eq.~\ref{eq:LamDec} to be of the form $\lambda_c(\Lambda,r_c)$. Using the equations from~\cite{gasbarri21_prosp_near_field_inter_tests_collap_model} to be,
\begin{equation}
	\lambda_c = \frac{-\Lambda (\kappa d)^2}{C \int e^{-\alpha^2} j_1(\frac{\alpha R}{r_c})^2 d\alpha \left(f(x)-1\right)}
	\label{eq:lamcLambda}
\end{equation}
where $C = 36 \sqrt{2/\pi} (M/m_0)^2 (r_c/R)^2$.

CSL tracking a constant $\Lambda$ defines a line in $\vec{\theta}=[r_c,\lambda_c]$ space.  We find $\Lambda$ such that the integral $\int p(\vec{\theta}|\vec{x})\,d\vec{\theta}$ below this line is $95\%$.  This is therefore a one-parameter problem which we solve by first scaling $\Lambda$ to find upper and lower bounds and then using iterative bisection to find the value of $\Lambda$ that gives us our exclusion line.

\section{Decoherence} \label{Sup:Deco}

\begin{figure}
	\centering
	\includegraphics[width=\linewidth]{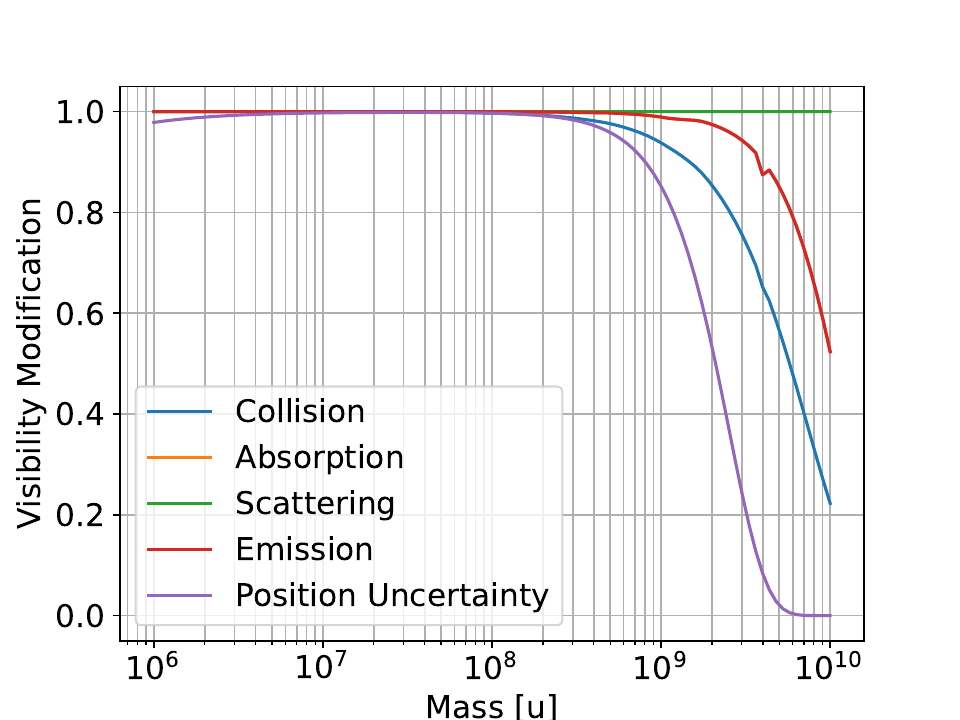}
	\caption{Reduction to the interference pattern visibility as a result of the various sources of environmental decoherence we consider using the MAQRO like environmental parameters.}
	\label{fig:DecoSources}
\end{figure}

\subsection{Position resolution} \label{Sup:Pos_Res}

Readout uncertainty is modelled as a constant offset, equal to the position uncertainty with which the initial state can be prepared, plus a constant rate of increase informed by \cite{esa2018qppf}: $\sigma_m(t) = \sigma_x + \left(10\,\text{nm}/100\,\text{s}\right)\, t$. 
By adjusting the rate at which the uncertainty increases, we can observe the effect this has on the information we expect to gain under the assumption that there is no CSL effect as shown Fig.~7.

\subsection{CSL for an extended particle}\label{Sup:CSL}

From reference~\cite[Eq.~6]{gasbarri21_prosp_near_field_inter_tests_collap_model} we obtain rate $\Gamma_\text{CSL}$ and resolution function $f_\text{CSL}$ for an extended particle~\footnote{There are minor differences from the reference: $\sqrt{2}$ in both and $\hbar x/q m_0^2$ in $f_\text{CSL}$.  There were identified as typos through private communication with Gasbarri.}
\begin{equation}
	\Gamma_\text{CSL}=\frac{4}{\sqrt{\pi}}\frac{\lambda_c r_c^3}{\hbar^3m_0^2} \int \ud q\, q^2 \exp\left(-r_c^2q^2/\hbar^2\right)\tilde{\mu}(q)^2
	\label{EQ:CSLGamma}
\end{equation}
\begin{multline}
f_\text{CSL}(x)=\frac{1}{\Gamma_\text{CSL}}
  \frac{4}{\sqrt{\pi}}\frac{\lambda_c r_c^3}{\hbar^3m_0^2}\\
  \int \ud q\, q^2 \exp\left(-r_c^2q^2/\hbar^2\right)\tilde{\mu}(q)^2\,\frac{\hbar}{x q}\,\Si\left(\frac{q x}{\hbar}\right)
  \label{EQ:CSLf(x)}
\end{multline}
where
\begin{equation}
\tilde{\mu}(\vec{q})=\int \ud \vec{x} \exp(-\vec{q}\cdot\vec{x}/\hbar) \mu(\vec{x})
\end{equation}
is the Fourier transform of the mass density function $\mu(\vec{x})$.  For the uniform density sphere, $\mu(\vec{x}) = \rho$ for $\lVert \vec{x}\rVert \leq R$, we use spherical symmetry to write $\tilde{\mu}(q)=\tilde{\mu}(\lVert\vec{q}\rVert)$. We find
\begin{equation}
\tilde{\mu}(q) = \frac{4\pi \hbar}{q}\rho R^2j_1\left(qR/\hbar\right).
\end{equation}
Hence, using $\alpha=q r_c/\hbar$ and $M=(4/3)\pi R^3 \rho$, we find
\begin{align}
  \Gamma_\text{CSL}
  &=A\,\lambda_c\,
  \int_0^\infty e^{-\alpha^2} j_1\left(\frac{\alpha R}{r_C}\right)^2\ud\alpha,\\
  f(x)
  &=\frac{A}{\Gamma_\text{CSL}}\,\lambda_c\,\frac{r_c}{x}
  \int_0^\infty e^{-\alpha^2} j_1\left(\frac{\alpha R}{r_C}\right)^2 \frac{1}{\alpha}\Si\left(\frac{\alpha x}{r_C}\right)\, \ud \alpha
\end{align}
with $A=\left(36/\sqrt{\pi}\right)\left(M/m_0\right)^2\left(r_c/R\right)^2$, as in the main text.

For a point-like particle $R\to 0$, these general results should reduce to point-like CSL: $\Gamma=(M/m_0)^2\lambda_c$ and $f_\text{CSL}(x)=\sqrt{\pi}(r_c/x)\erf\left(x/2r_c\right)$.  Each expression contains a term $j_1(ab)/b$ which becomes $b/3$ in the limit $a\to 0$.  Then, using the integral results 
\begin{align}
\int_0^\infty \ud\alpha^2\,e^{-\alpha}\alpha^2 &=\sqrt{\pi}/4 \\
\int_0^\infty \ud\alpha\,e^{-\alpha}\alpha  \Si(\alpha \beta)&=(\pi/4)\erf(\beta/2)
\end{align}
we find that the point-like CSL expressions are indeed recovered in this limit.

\subsection{Blackbody and collisional decoherence} \label{Sup:BB_Col_Deco}

This section summarizes the implementation used in this work, following reference~\cite{bateman2014nearfield}.

Blackbody radiation leads to decoherence through absorption, emission, and Rayleigh scattering.
Rates per unit wave-number $k$ at temperature $T$ are computed as 
\begin{equation}
\gamma(k,T) = \int_0^\infty \frac{(k/\pi)^2 \sigma(k)}{\exp{\left(\hbar ck/k_BT\right)}-1}
\end{equation}
where $\sigma_\text{abs}(k)=k\Im\left[\chi(k)\right]$ for absorption and emission and $\sigma_\text{sca}(k)=k^4\lvert\chi(k)\rvert^2/6\pi$ for Rayleigh scattering.  The susceptibility $\chi$ is found from material permittivity $\epsilon$ and the Clausius--Mossotti relation: $\chi(k)=3V\left[\epsilon(k)-1\right]/\left[\epsilon(k)+1\right]$ where $V$ is the particle volume; the particle is assumed sub-wavelength for all relevant blackbody wavelengths.

For simplicity, and confident that it gives an upper-estimate for decoherence and the difference is small, we assume the particle does not cool during free-flight.  Hence, equations for decoherence for emission and absorption of blackbody radiation are the same, with the only difference being temperature.

To compute the corresponding $R_n$ decoherence Talbot coefficient requires total rate $\Gamma$ and function $f$.  These are $\Gamma = \int_0^\infty \gamma(k,T)\,\ud k$ for each of the mechanisms,
\begin{equation}
f_\text{abs}(x) = c\int_0^\infty \frac{\gamma_\text{abs}}{\Gamma_\text{abs}}\frac{\Si(kx)}{kx}\,\ud k
\end{equation}
for absorption (similarly for emission), and
\begin{equation}
f_\text{sca}(x)=c\int_0^\infty \frac{\gamma_\text{sca}(k,T)}{\Gamma_\text{sca}}\left[\frac{\Si(2kx)}{kx}-\sinc^2(kx)\right]\,\ud k
\end{equation}
for Rayleigh scattering.

We treat collisional events with background gas as resolving position far better than the grating wavelength and hence we only require the rate $\Gamma_\text{col}$; this is computed as in reference \cite{bateman2014nearfield}.

\begin{figure}
	\centering
	\includegraphics[width = \linewidth]{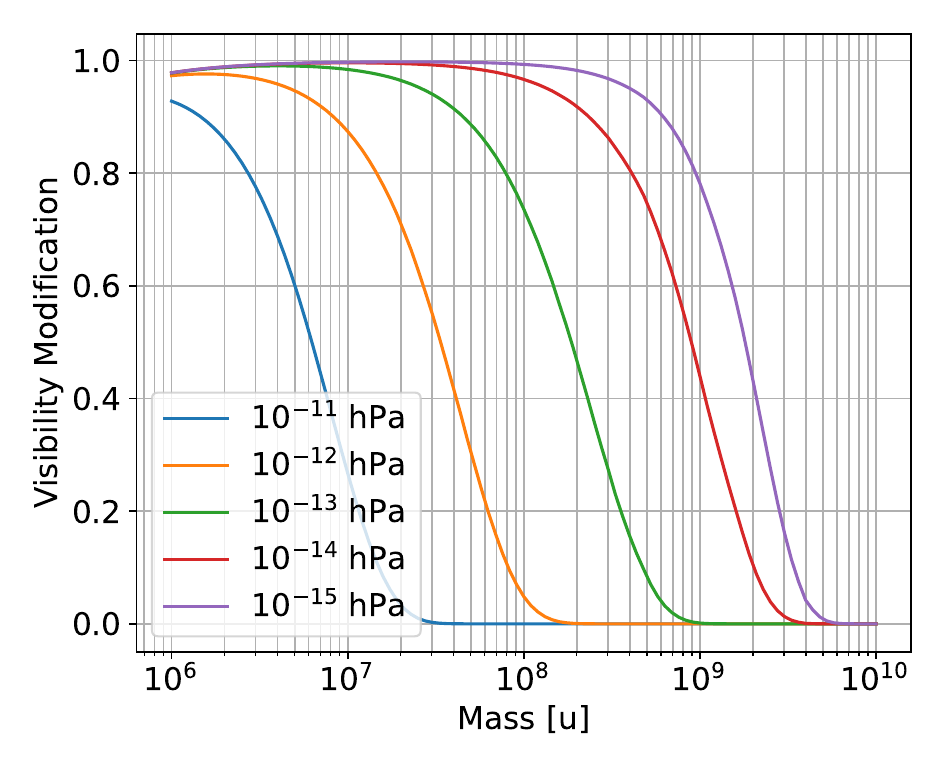}
	\caption{Total effect of environmental decoherence on the fringe visibilities, $R_1^\text{oth}$, for various experimental scenarios with varying residual gas pressure.}
	\label{fig:TotDeco}
\end{figure}

\section{Talbot Coefficients for a Mie Particle} \label{Sup:Talbot}

We derive the scattering decoherence on a particle in the standing wave grating using the method described in \cite{pflanzer_master-equation_2012}. In contrast with previous work \cite{belenchia2019talbot-lau}, we find that scattering into $\theta$ and $\phi$ polarizations must be treated as separate processes.

The electromagnetic field that forms the bath is given by
\begin{multline}
	\vec{\hat{E}}_B(\hat{\vec{r}}) = i \int_{\vec{k}}\sum_{\nu} \frac{\hbar \omega_{\vec{k}}}{2V_q \epsilon_0} \sqrt{\frac{\hbar \omega_{\vec{k}}}{2\epsilon_0}}\\
	 \left(\vec{\varepsilon }_{\vec{k},\nu}\hat{a}_{\vec{k},\nu}e^{i\vec{k}\cdot\hat{\vec{r}}} - \vec{\varepsilon }^*_{\vec{k},\nu}\hat{a}^{\dagger}_{\vec{k},\nu}e^{-i\vec{k}\cdot\hat{\vec{r}}}\right) 
\end{multline}
where $\hat{a}_{\vec{k}\nu}$ ($\hat{a}^{\dagger}_{\vec{k},\nu}$) is the annihilation (creation) operator, $\vec{\varepsilon}_{\vec{k},\nu}$ is the polarization vector, $\vec{k}$ is the wave vector, $\nu$ denotes the two independent polarization directions $\theta$ and $\phi$, $w_k=ck$, and $k=|\vec{k}|$. We also define $V_q$ as the quantization volume defined by the boundaries of the experiment given by $V_q = L^3$ where $L$ is the length of a box.
\par
We are then able to perform the calculations given in~\cite{pflanzer_master-equation_2012} whilst maintaining the explicit dependence on the phase to obtain the transition matrix,
\begin{equation}
	\mathcal{T}_{\vec{k}\nu,c}(\vec{\hat{r}}) = \sum_{\nu}\int d\vec{k} \bra{\vec{c}}\ket{\vec{k'},\nu'} \mathcal{T}^*_{\vec{k'}\nu',\vec{k}\nu}(\vec{\hat{r}})
\end{equation}
and the scattering superoperator,
\begin{multline}
	\mathcal{L}(\rho) = |\alpha|^2 \sum_{\nu} \int d\vec{k'}\delta(\omega_k - \omega_0)\\
	 \left(2\mathcal{T}_{\vec{k}\nu,c}(\hat{r}) \rho \mathcal{T}^*_{c,\vec{k}\nu}(\hat{r}) - \left\{|\mathcal{T}_{\vec{k}\nu,c}(\hat{r})|^2,\rho\right\}\right).
\end{multline}

For a standing wave with linear polarization, the mode function $\ket{\vec{c}}$ can be described by,
\begin{multline}
	\bra{\vec{k},\nu}\ket{\vec{c}} = \frac{1}{V_0} \int d\vec{x} e^{-i\vec{k}\cdot\vec{x}} \vec{\varepsilon}_{\vec{k},\nu}\cdot\vec{\varepsilon}_{\vec{d}}f(y,z)\cos(k_0x)\\
	 \approx \frac{\omega_0 \vec{\varepsilon}_{k,\nu}\cdot\vec{\varepsilon}_{k_z,\nu'}}{\sqrt{V_0}} \tilde{f}(k_y,k_z)\delta(k_y)\delta(k_z)\delta(k_x-\omega_0^2)
\end{multline}
where we have made the assumption that the laser field has a very large spot area. From this we can obtain,
\begin{multline}
	\mathcal{T}_{\vec{k}\nu,\vec{c}}(\vec{\hat{r}})\approx \frac{1}{4\pi \omega_k \sqrt{V_0}}\\
	 \left(e^{-i\vec{k_0}\cdot\vec{\hat{r}}} f^*_{\nu_0,\nu}(\vec{k_0},\vec{k}) + e^{i\vec{k_0}\cdot\vec{\hat{r}}} f^*_{\nu_0,\nu}(-\vec{k_0},\vec{k})\right).
\end{multline}

If we assume that the wave is polarized in the x direction, we recover the vector scattering amplitude,
\begin{equation}
	\sum_{\nu}\hat{e}_{\nu} f^*_{\nu}(\vec{k_0},\vec{k}) = (S_2(\theta)\cos\phi \hat{e}_{\theta} -  S_1(\theta)\sin\phi \hat{e}_{\phi}).
\end{equation}
Note that $\hat{e}_{\theta}$ and $\hat{e}_{\phi}$ are orthogonal components of the scattered field polarization such that $\hat{e}_{\theta}\cdot\hat{e}_{\phi}=0$. This allows us to rewrite the scattering superoperator as
\begin{equation}
	\begin{split}
		\mathcal{L}(\rho) = |\alpha|^2 \int d\vec{k'}&\delta(\omega_k - \omega_0)\\
		 &\left(2\mathcal{T}_{\vec{k}\phi,c}(\hat{r}) \rho \mathcal{T}^*_{c,\vec{k}\phi}(\hat{r}) - \left\{|\mathcal{T}_{\vec{k}\phi,c}(\hat{r})|^2,\rho\right\}\right) \\
		+ |\alpha|^2 \int d\vec{k'}&\delta(\omega_k - \omega_0)\\
		 &\left(2\mathcal{T}_{\vec{k}\theta,c}(\hat{r}) \rho \mathcal{T}^*_{c,\vec{k}\theta}(\hat{r}) - \left\{|\mathcal{T}_{\vec{k}\theta,c}(\hat{r})|^2,\rho\right\}\right).
	\end{split}
\end{equation}

Through lengthy algebra we can obtain the scattering mask,

\begin{equation}
	R_{sca} = \exp\left\{\sum_{\nu}\left[F_{\nu}(s) + a_{\nu}(s)\cos(2kx) + ib_{\nu}(s)\sin(2kx) \right]\right\}
\end{equation}
where
\begin{equation}
	\begin{split}
		a_{\nu} &= \frac{8\pi}{\hbar c k}\frac{E_G}{a_G} \int d\Omega \Re(f^*_{\nu}(\vec{k},\vec{k'})f_{\nu}(\vec{k},\vec{k'})[\cos(kn_x s)-\cos(ks)]\\
		b_{\nu} &= \frac{8\pi}{\hbar c k}\frac{E_G}{a_G} \int d\Omega \Im(f^*_{\nu}(\vec{k},\vec{k'})f_{\nu}(\vec{k},\vec{k'})\sin(kn_x s)\\
		F_{\nu} &= \frac{8\pi}{\hbar c k}\frac{E_G}{a_G} \int d\Omega |f_{\nu}(\vec{k},\vec{k'})|^2 [\cos(k(1-n_x)s) -1].\\
	\end{split}
\end{equation}
We can then combine this with the absorption mask, $R_\text{abs}(x,x') = \exp \left[-2n_0\sin^2\left(k_0\frac{x+x'}{2}\right) \sin^2\left(k_0\frac{x-x'}{2}\right)\right]$ where $n_0 = \frac{I_0}{cF_0}\sigma_\text{abs}\phi_0$ is the mean number of absorbed photons and we have used the absorption cross section $\sigma_\text{abs}$ given by Mie theory.
\par
We take the Fourier coefficients of this total decoherence mask and, with use of Graf's addition theorem, convolve it with the coherent grating effects as described in~\cite{belenchia2019talbot-lau} to obtain the final Talbot coefficients as,
\begin{widetext}
\begin{multline}
	\tilde{B}_n\left(\frac{s}{d}\right) = e^{F_{\phi}(s)+F_{\theta}(s)-\zeta_\text{abs}}
	\sum_{k=-\infty}^{\infty} \left[\frac{\zeta_\text{coh} + a_{\phi}(s)+a_{\theta}(s) + \zeta_{abs}}{\zeta_\text{coh} - a_{\phi}(s)-a_{\theta}(s) - \zeta_\text{abs}}\right] ^{\frac{n+k}{2}}\\
	 J_k[b_{\phi}(s)+b_{\theta}(s)] J_{n+k}\left[\sign(\zeta_\text{coh} - a_{\phi}(s)-a_{\theta}(s) - \zeta_\text{abs}) \sqrt{\zeta^2_\text{coh} -(a_{\phi}(s)+a_{\theta}(s) + \zeta_\text{abs})^2}\right]
\end{multline}
\end{widetext}
where we have made the substitutions 
\begin{align}
	\zeta_\text{abs} &= \frac{n_0}{2}(1-\cos(\frac{\pi s}{d}))\\
	\zeta_\text{coh} &= \phi_0\sin(\frac{\pi s}{d})
\end{align}

\end{document}